# Measuring the electron neutrino mass using the electron capture decay of $^{163}$Ho

Joel Ullom (joel.ullom@nist.gov), Daniel Schmidt, NIST
Simon Bandler, Thomas Stevenson, NASA Goddard Space Flight Center
Mark Croce, Los Alamos National Laboratory
Katrina Koehler, Houghton College
Matteo De Gerone, INFN Genoa
Loredana Gastaldo, Christian Enss, Heidelberg University
Geonbo Kim, Lawrence Livermore National Laboratory
Angelo Nucciotti, Stefano Ragazzi, University of Milano-Bicocca and INFN Milano
Kyle Leach, Colorado School of Mines
Diana Parno, Carnegie Mellon University
Brian Mong, Josef Frisch, Christopher Kenney, SLAC National Laboratory

While the mass differences between neutrino mass states are known, their absolute masses and mass hierarchy have not yet been determined. Determining the mass of neutrinos provides access to physics beyond the Standard Model (SM) and the resulting value has implications for the growth of large-scale structure in the universe over cosmic history. Because of the importance of the topic, a number of efforts are already underway to determine the mass of neutrinos including direct kinematic measurements and indirect measurements of astrophysical phenomena that constrain the sum of the mass eigenstates through models of cosmic evolution. Here, we advocate for a collaborative international effort to perform a kinematic determination of the effective electron neutrino mass using calorimetric measurements of the decay of $^{163}$Ho. This effort is justified by the success of current experiments using the technique, its high benefit-to-cost ratio, the value of approaches with different systematic errors, and the value of measuring the electron neutrino mass ($m_{\nu e}$) rather than the electron anti-neutrino mass.

The electron capture decay of $^{163}$Ho provides an attractive system for kinematic measurements of the effective electron neutrino mass [De Rujula, 1983]. When $^{163}$Ho is embedded in a calorimetric sensor, each decay deposits energy in the sensor equal to the Q-value of the reaction minus the energy of the departing neutrino. The rest mass of the neutrino is manifested as a deficit of events in the region of the decay spectrum near the Q-value. $^{163}$Ho is particularly attractive because of its low Q-value of 2.833 keV [Eliseev, 2015] and its reasonable half-life of 4570 years [Baisden, 1983]. A low Q-value and the proximity to the highest resonance related to the capture of 3s electrons increases the fraction of events in the interesting endpoint region of the spectrum and a short half-life reduces the amount of Ho that must be embedded in a sensor to achieve a target count rate. Measuring keV-scale energy depositions with eV-scale accuracy is a task that is already achieved by modern cryogenic microcalorimeters [Kempf, 2018; Smith, 2012]. Both transition-edge sensors (TESs) and magnetic microcalorimeters (MMCs) are attractive candidates, and other viable calorimetric sensors may yet emerge. A large array of microcalorimeters each with embedded $^{163}$Ho atoms can produce a high statistics decay spectrum whose analysis provides sub-eV sensitivity to $m_{\nu e}$ [Gastaldo, 2017; Nucciotti, 2014]. This approach is the subject of our white paper.

The state-of-the field for neutrino mass measurements is complex but also fertile for new approaches.  After a promising start, KATRIN has set a 0.8 eV upper bound on the effective electron anti-neutrino mass and is targeting a kinematic measurement based on tritium β-decay with 0.2 eV sensitivity within five years [Aker, 2022].  The Project 8 team is interested in using cyclotron resonance emission spectroscopy to determine the effective electron anti-neutrino mass but more steps have still to be done to demonstrate scalability to the necessary experimental sensitivity [Esfahani, 2017].  Measurements of astrophysical phenomena constrain the sum of the masses of the three neutrino flavors to < 0.12 eV [Planck, 2018] in the context of the ΛCDM model and the summed mass sensitivities of future experiments such as Simons Observatory are predicted to be as good as 0.02 eV [Ade, 2019].  Astrophysical measurements, however, are less attractive from a fundamental discovery standpoint as they are heavily model dependent.

Surveying this experimental landscape, there is a need for an alternative kinematic technique with mass sensitivity comparable to or better than KATRIN's projected limits and with very different systematic error terms from KATRIN or Project 8.  In a corroboratory role, a new technique could resolve disagreement between KATRIN and astrophysical results.  If its sensitivity surpassed KATRIN's, then a new technique could provide the most stringent kinematic measurement of the neutrino mass to date.  Additionally, sensitivity to normal-matter mass states, rather than the anti-neutrino mass states in $^3$H, could potentially provide additional constraints on possible CPT violating theories.

Many of the risks associated with the $^{163}$Ho approach have already been retired. The ECHo collaboration has demonstrated decay spectra with 275,000 counts using a small array of magnetic microcalorimeters [Velte, 2019] and is presently analyzing data corresponding to a spectrum with ~$10^8$ counts.  The HOLMES collaboration is developing multiplexed transition-edge sensors (TESs) for this purpose [Faverzani, 2020].  The TESs and readout SQUIDs for HOLMES are provided by NIST, so the U.S. already has a presence in this field.  An exploratory project at Los Alamos measured $^{163}$Ho spectra with NIST TESs [Croce, 2016].  Together, these three efforts have demonstrated the ability to synthesize, purify, and embed $^{163}$Ho, to fabricate microcalorimeters, to measure $^{163}$Ho decay spectra at application-relevant resolution levels, and to use modern multiplexing techniques to read out arrays of microcalorimeters.  In addition, recent theoretical work has retired risk about the shape of the $^{163}$Ho spectral endpoint [Brass, 2020 and references therein].

To achieve sensitivity to neutrino masses near 0.2 eV requires the measurement of ≳ $10^{16}$ decays and, consequently, the realization of a large number of microcalorimeter channels.  To measure $10^{16}$ decays in 5 years will require about 64 MBq of embedded $^{163}$Ho. The mass of $^{163}$Ho that corresponds to 64 MBq of activity is 3.55 mg (or 1.3x$10^{19}$ atoms) which is quite a large quantity for an artificial radioisotope.  However, production on this scale is feasible given the amounts which have been produced at the Institut Laue-Langevin high flux reactor for ECHo and HOLMES [Dorrer, 2018; Heinitz, 2018].  Choosing an activity per sensor of 64 Bq, then $10^6$ sensors are required.  Clearly, the development of $10^6$ sensors with embedded $^{163}$Ho will be a challenge, but a comparable number (5x$10^5$) of cryogenic detectors are already planned for the DOE/NSF CMB-S4 program to study the cosmic microwave background. Further, the physical scale of $^{163}$Ho experiments is modest compared to some other approaches.  Each sensor and its readout circuitry will likely occupy < 5 mm$^2$ of area on a 350 µm thick silicon substrate so the whole experiment will fit on < 5 m$^2$ of planar Si weighing about 4.3 kg.  Sensitivity to neutrino mass is believed to scale as ~1/counts$^{1/4}$ [Nucciotti, 2014], so to achieve 0.1 eV sensitivity and surpass the anticipated results of KATRIN, for example, will require 1.6x10$^{17}$ decays.  It is an open question whether increases in sensitivity should be achieved with more sensors, more $^{163}$Ho per sensor, or more subtle improvements such as better timing resolution to suppress backgrounds.

An international collaboration between groups in the US and Europe is an attractive path to execute a large-scale $^{163}$Ho experiment.  The $^{163}$Ho approach does not depend on a unique apparatus such as a large electromagnetic spectrometer.  Instead, the sensors could be distributed among several US and European sites.  Furthermore, the number of sensors could be increased in a series of stages allowing science to start early in the program with the data collection rate growing as more sensors are produced and deployed.  In order to maximize efficiency, different institutes would provide different subsystems that reflect their particular strengths.  While R&D on microcalorimeter-based $^{163}$Ho neutrino experiments is presently more mature in Europe than in the U.S., for example see [Mantegazzini, 2022], the U.S. community can make important contributions.  Work in the U.S. on both the cold [Mates, 2017] and warm components [Henderson, 2018] needed for multiplexed readout is world-leading.  In addition, the U.S. has excellent facilities for fabricating cryogenic detectors and SQUID multiplexers.

Before beginning an experiment with large numbers of microcalorimeters, there is further exploratory work that should be performed. Important tasks that could be constructively pursued at the few-sensor scale with modest levels of funding include:

1. <u>Optimization of $^{163}$Ho embedding</u>. This task includes determining the optimal $^{163}$Ho loading per sensor as well as the optimal embedding mechanism.  The parameter space is complex and interesting.  For example, the achievable Ho concentration depends on the embedding mechanism, the heat capacity contribution per Ho atom depends on the concentration and temperature [Herbst, 2021], and the speed and energy resolution of the sensors depends on the total sensor heat capacity, including the holmium contribution.  This task should also include assessment of the impact, if any, of the chemical environment of the holmium on the decay spectrum.
2. <u>Demonstration of high yield detector fabrication including the embedding of $^{163}$Ho</u>. Some embedding schemes such as ion implantation likely scale well to the use of detector arrays.  For other schemes, such as embedding from liquid solution, further research is required.  Regardless of the selected path, a successful demonstration of array-scale embedding is needed.
3. <u>Design of a cost-effective readout scheme</u>. Microcalorimeter readout techniques based on GHz-domain frequency-division multiplexing provide a clear technical path to instruments with 10$^6$ microcalorimeters.  However, it remains to be seen what will be the readout cost per channel: likely values range between $1 and $100 per channel.  Readout costs at the upper end of this range could be a significant driver of the overall cost, channel count, and science reach of a large $^{163}$Ho experiment, and costs at the lower end of the range are subdominant to other program components.  This is a time of intense development in the field of microwave electronics because of the growth in 5G telecommunications, for example with emerging RFSoCs likely to displace instruments based on discrete ADC, DAC, and FPGA chips.  A careful study of readout requirements and electronics solutions that leverage commercial components could greatly reduce the cost a future large-scale program.
4. <u>Further studies of neutrino mass sensitivity optimization</u>.  Multiple factors affect the ultimate sensitivity of $^{163}$Ho experiments including the energy resolution of the sensors, the number of sensors, the count rate per sensor, and the effectiveness with which sources of background including coincident events can be rejected.  A combination of simulation and small-scale experimental studies could produce a design-of-experiment that yields major dividends in the ultimate sensitivity of a future large-scale program.
5. <u>Demonstration of real-time signal processing</u>.  While off-line storage and analysis of $\gtrsim 10^{16}$ event records is possible, the attractions of real-time event processing and analysis are obvious.  Work on appropriate algorithms can begin now with small numbers of sensors, or even a single sensor.

The steps above can be performed in collaboration with the ECHo and HOLMES teams to accelerate progress.  We anticipate that after roughly three years of development there will be a solid foundation from which to launch an impactful $^{163}$Ho experiment that grows through a series of stages to reach the scale of $10^6$ sensors or more.

In summary, studies of $^{163}$Ho can provide a direct measurement of the neutrino mass scale with competitive sensitivity and very different systematic errors compared to other approaches.  Such a result will be extremely valuable given the current landscape of neutrino mass efforts.  An international effort that leverages the strengths of institutes around the world can accomplish this goal in a cost-effective manner.  This effort will both benefit from and accelerate other cryogenic detector projects such as CMB-S4, successors to CUORE, and the development of x-ray spectrometers for DOE light sources.

**References**

[Ade, 2019] Peter Ade et al, J. Cosmology and Astroparticle Physics 02 (2019) 056
[Aker, 2019] M. Aker et al, Nature Physics 18 (2022) 160
[Baisden, 1983] P. A. Baisden et al, Phys. Rev. C 28 (1983) 337
[Brass, 2020] M. Brass and M. W. Haverkort, New J. Phys. 22 (2020) 093018
[Croce, 2016] M. P. Croce et al, J. Low Temp. Phys. 184 (2016) 958
[De Rujula, 1983] A. De Rujula and M. Lusignoli, Phys. Lett. B 118 (1983) 429
[Dorrer, 2018] H. Dorrer et al, Radiochimica Acta 106/7 (2018) 535
[Eliseev, 2015] S. Eliseev et al, PRL 115 (2015) 062501
[Esfahani, 2017] A. A. Esfahani et al, J. Phys. G: Nucl. Part. Phys. 44 (2017) 054004
[Faverzani, 2020] M. Faverzani et al, J. Low Temp. Phys. 199 (2020) 1098
[Gastaldo, 2017] L. Gastaldo et al, Eur. Phys. J. Special Topics 226 (2017) 1623
[Heinitz, 2018] S. Heinitz et al, PLoS ONE 13(8) (2018) e0200910
[Henderson, 2018] S. W. Henderson et al, Proc. SPIE 10708 Millimeter, Submillimeter, and Far-Infrared Detectors and Instrumentation for Astronomy IX (2018) 1070819
[Herbst, 2021] M. Herbst et al, J. Low Temp. Phys. 202 (2021) 106
[Kempf, 2018] S. Kempf et al, J. Low Temp. Phys. 193 (2018) 365
[Mates, 2017] J. A. B. Mates et al, Appl. Phys. Lett. 111 (2017) 062601
[Mantegazzini, 2022] F. Mantegazzini et al, NIM A 1030 (2022) 166406
[Nucciotti, 2014] A. Nucciotti, Eur. Phys. J. C 74 (2014) 3161
[Planck, 2018] Planck 2018 results. VI. Cosmological parameters, arXiv:1807.06209
[Smith, 2012] S. Smith et al, J. Low Temp. Phys. 167 (2012) 168
[Velte, 2019] C. Velte et al, Eur. Phys. J. C 79 (2019) 1026

,